\documentclass[useAMS,usenatbib,letterpaper]{mn2e}
\usepackage{amsmath,amssymb}
\usepackage{graphicx}

\def\Omm{{\Omega_m}}
\def\Ommz{{\Omega_m^{\,z}}}

\def\Omk{{\Omega_k}}
\def\Oml{{\Omega_{\Lambda}}}

\def\apj{ApJ}

\def\apjl{ApJ}
\def\mnras{MNRAS}

\def\aj{AJ}

\def\nat{Nat}

\def\apjs{ApJS}

\newcommand{\beq}{
\begin{equation}
}
\newcommand{\eeq}{
\end{equation}
}

\def\simlt{\mathrel{\rlap{\lower 3pt\hbox{$\sim$}}\raise 2.0pt\hbox{$<$}}}
\def\simgt{\mathrel{\rlap{\lower 3pt\hbox{$\sim$}} \raise 2.0pt\hbox{$>$}}}

\def\gsim{ \lower .75ex \hbox{$\sim$} \llap{\raise .27ex \hbox{$>$}} }
\def\lsim{ \lower .75ex\hbox{$\sim$} \llap{\raise .27ex \hbox{$<$}} }

\def\kmps{{\rm\thinspace km \thinspace s^{-1}}}

\def\beq{\begin{equation}}
\def\eeq{\end{equation}}

\def\Omm{{\Omega_m}}
\def\Ommz{{\Omega_m^{\,z}}}

\def\Omk{{\Omega_k}}
\def\Oml{{\Omega_{\Lambda}}}

\title[Blazars in the early Universe]
{Blazars in the early Universe}
\author[Volonteri et al.]
{M. Volonteri$^1$,\thanks{Email: martav@umich.edu}
F. Haardt$^{2,3}$, G. Ghisellini$^4$, R. Della Ceca$^5$
\\
\\
$^1$Astronomy Department, University of Michigan, Ann Arbor, MI 48109, USA \\
$^2$Universit\`a dell'Insubria, Dipartimento di Fisica e Matematica, Via Valleggio 11, I--22100 Como, Italy;\\
$^3$INFN, Sezione di Milano-Bicocca, 20126 Milano, Italy\\
$^4$INAF -- Osservatorio Astronomico di Brera, Via Bianchi 46, I--23807 Merate, Italy\\
$^5$INAF -- Osservatorio Astronomico di Brera, Via Brera 28, I--20100 Milano, Italy\\
}

\begin{document}  

\maketitle

\begin{abstract}
We investigate the relative occurrence of radio--loud and radio-quiet quasars
in the first billion years of the Universe, powered by black holes heavier than one billion solar masses.
We consider the 
sample of high--redshfit blazars detected in the hard X--ray 
band in the 3--years all sky survey performed by the Burst Alert Telescope 
(BAT) onboard the {\it Swift} satellite. All the black holes powering these blazars exceed a billion
solar mass, with accretion luminosities close to the Eddington limit.
For each blazar pointing at us, there must be hundreds of similar sources (having 
black holes of similar masses) pointing elsewhere. 
This puts constraints on the density of billion solar masses black holes at high redshift 
($z>4$), and on the relative importance of (jetted) radio--loud vs radio--quiet sources. 
We compare the expected number of high redshift radio--loud sources
with the high luminosity radio--loud quasars detected in the Sloan Digital Sky Survey
(SDSS), finding agreement up to $z\sim 3$, but a serious deficit at $z>3$
of SDSS radio--loud quasars with respect to the expectations. 
We suggest that the most likely explanations for this disagreement are:  
i) the ratio of blazar to misaligned radio--sources decreases by an 
order of magnitude above $z=3$, possibly as a result of a decrease of 
the average bulk Lorentz factor;
ii) the SDSS misses a large fraction of radio--loud sources at high redshifts,
iii) the SDSS misses {\it both} radio--loud and radio--quiet quasars
at high redshift, possibly because of obscuration or because of collimation of the
optical--UV continuum in systems accreting near Eddington.
These explanations imply very different number density of heavy black holes
at high redshifts, that we discuss in the framework of the current ideas about 
the relations of dark matter haloes at high redshifts and the black hole they host.
\end{abstract}

\section{Introduction}

\begin{figure}
\vskip -0.4 cm
\includegraphics[width= \columnwidth]{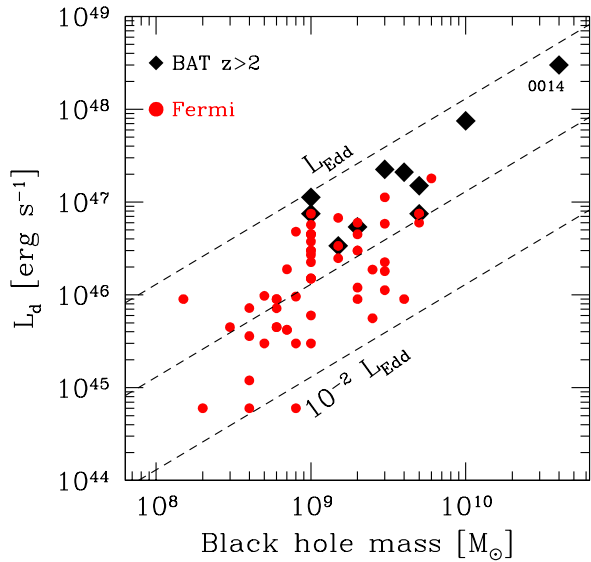}
\vskip -0.5 cm
\caption{
Accretion disk luminosity $L_{\rm d}$
as a function of black hole mass for blazars with $z>2$ in the
BAT sample (diamonds; A09) and in the 1LAC {\it Fermi}/LAT sample
(circles, Abdo et al. 2010), as studied in G10 and in Ghisellini et al. (2011).
The high redshift BAT blazars are all characterized by black holes
with $M>10^9 M_\odot$ and by $L_{\rm d}/L_{\rm Edd}>0.1$.
}
\label{blazars}
\end{figure}

Ajello et al. (2009, hereafter A09) recently published the list 
of blazars detected in the all sky survey by {\it Swift}/BAT, 
between March 2005 and March 2008.
BAT is a coded mask telescope designed to detect Gamma Ray Bursts (GRBs),
has a large field of view ($120^\circ\times 90^\circ$, partially coded) and
is sensitive in the [15--150 keV] energy range.
This instrument was specifically designed to detect GRBs, but since 
GRBs are distributed isotropically in the sky, BAT performed, 
as a by product, an all sky survey with a reasonably 
uniform sky coverage, at a limiting sensitivity of the order 
of 1 mCrab in the 15--55 keV range 
(equivalent to $1.27\times 10^{-11}$ erg cm$^{-2}$ s$^{-1}$) 
in 1 Ms exposure, see A09).
Taking the period March 2005 -- March 2008, and evaluating the image
resulting from the superposition of all observations in this period,
BAT detected 38 blazars (A09), of which 26 are Flat Spectrum Radio Quasars
(FSRQs) and 12 are BL Lac objects, once the Galactic plane ($|b<15^\circ|$) 
is excluded from the analysis.
A09 reported an average exposure of 4.3 Ms, and considered the [15--55 keV] 
energy range, to avoid background problems at higher energies.
The well defined sky coverage and sources selection criteria
makes the list of the found blazars a complete, flux limited, 
sample, that enabled A09 to calculate the luminosity function (LF)
and the possible cosmic evolutions of FSRQs and BL Lacs,
together with their contribution to the hard X--ray background.
A09 also stressed the fact that the detected BAT blazars at high redshift
are among the most powerful blazars and could be associated with powerful
accreting systems. 
Within the BAT sample, there are 10 blazars (all FSRQs) at redshift greater
than 2, and 5 at redshift between 3 and 4.
All (and only) these blazars have a X--ray luminosity exceeding $L_X=2\times 10^{47}$ erg s$^{-1}$.
All these sources have been studied by Ghisellini et al. (2010, hereafter G10),
that showed that their optical--UV emission is dominated by the emission 
of their accretion disk, with no contamination from the beamed non--thermal
continuum, even if the latter is dominating the total bolometric luminosity.
Fitting the optical--UV emission with a standard Shakura \& Sunyaev (1973) 
accretion disk, it was possible to estimate both the mass and the disk
luminosity.
These high redshift blazars are shown in Fig.~\ref{blazars} (diamonds) together with all the blazars
with $z>2$ detected above 100 MeV in the 11 months all sky survey performed by {\it Fermi}/LAT
(Abdo et al. 2010) and studied in Ghisellini et al. (2011). 
Fig. \ref{blazars} shows that all high redshift BAT blazars are 
characterized by large Eddington ratio of order 0.2--1
and by black holes heavier than $10^9 M_\odot$

Since these objects are at high redshifts, our finding has
important implications on the number density of heavy black holes,
especially if we consider that for each blazar pointing at us, there
must be hundreds of similar sources (having black holes of similar masses)
pointing elsewhere.
In fact, if the emitting plasma is moving with a bulk Lorentz factor $\Gamma$ in one
direction, the number of sources observed within the beaming angle $1/\Gamma$ is 
only a fraction $1/(2\Gamma^2)$ of the the sources pointing in other directions.

Taking the luminosity function of A09 at face value and calculating the expected 
number of luminous sources (i.e. $L_x>2\times 10^{47}$ erg s$^{-1}$, likely hosting a black hole with 
$M>M_9$, where $M_9=10^9 M_\odot$) at $z>4$, one finds that the number density of heavy black holes 
of jetted sources is close to or even greater than the upper limit defined by 
standard ``dark matter halo--black holes" relationships at the largest redshifts. 
G10 then corrected the original A09 luminosity function by 
assuming an evolutionary model that is equal to the A09 one up to $z\sim 4.3$ 
(where they measure the  peak of the density of high X--ray luminosity blazars), 
and then cuts off exponentially (``minimal" LF).
This ``minimal LF" was consistent with the constraints posed by standard
``dark matter halo--black holes" relationships and consistent with the constraints
given by the existence of a few blazars (discovered serendipitously) at $z>4$ (see below).

In this paper we explore the implications of G10 results in view of the properties of the 
radio quiet and radio--loud populations, their redshift evolution 
and their connection to host dark matter halos.
First we check if the expected number of radio--loud sources calculated from the
A09 and G10 luminosity functions agrees with those detected in the SDSS survey of quasars.
As discussed in Section 2, while a rough agreement is found up to $z\sim3$, there is a serious 
deficit of SDSS radio--loud sources above this redshift.
We then investigate the possible reasons of this discrepancy, discussing
three possible solutions.
While we cannot confidently select one of these, we point out the consequences
they have on our understanding of the physics of jets and on the relationship between dark matter
halos and the mass of the black holes they host.

\begin{figure}
\vskip -0.2 cm
\includegraphics[width= \columnwidth]{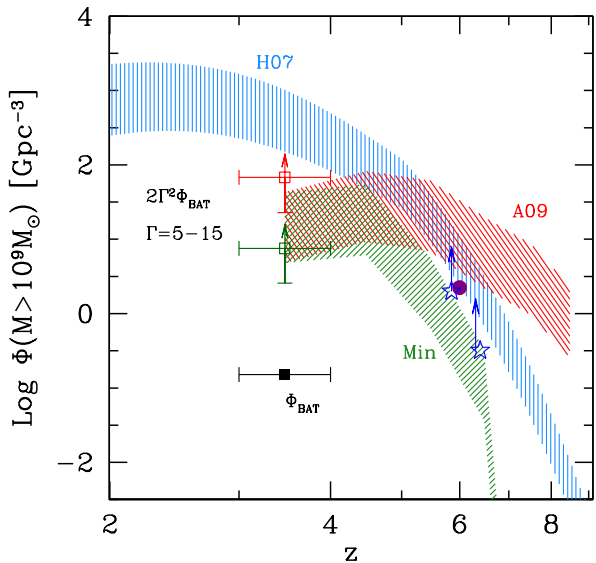}
\vskip -0.7 cm
\caption{
Number density of black holes with $M>10^9M_\odot$ as a function of redshift. 
The filled black square in the  $3<z<4$ bin is taken directly from Fig. 10 of A09. 
The two empty squares account for the population of misaligned sources,
multiplying by $2\Gamma^2$, with $\Gamma=5$ and $\Gamma=15$. 
Red hatched area: number density of heavy black holes in radio--loud sources derived from the blazar LF
studied in A09, assuming $\Gamma=5$ (lower bound), or $\Gamma=15$ (upper bound). 
Green hatched area: number density of heavy black holes assuming the ``minimal"  cosmic evolution 
of the LF of blazars (G10), with $\Gamma=5$ (lower bound), or $\Gamma=15$ (upper bound).
Blue hatched area:  number density of heavy black holes in radio--quiet quasars 
(assuming the LF and its evolution of H07), assuming
an average Eddington fraction of $f_{\rm Edd}=0.3$ (upper boundary) or
$f_{\rm Edd}=1$ (lower boundary). 
Blue stars: number density of $M>10^9M_\odot$ derived from the existence of the black 
holes analyzed in Kurk et al. (2007). 
Purple dot: number density of $M>10^9M_\odot$ derived from the mass function of black holes 
at $z=6$ proposed by Willott et al. (2010).
}
\label{data}
\end{figure}

\section{Radio--loud high redshift sources}

One can estimate the volume density of high redshift blazars hosting a 
black hole of mass larger than $M_9$ using the cosmological 
evolution model of A09 along with its high--$z$ cut--off (i.e. ``minimal")
version, assuming that all blazars with $L_X>2\times 10^{47}$
erg s$^{-1}$ have a $M>M_9$ black hole (G10).
We cannot exclude that blazars with lower X--ray luminosity also host
massive black holes, so the ``observational" points, strictly
speaking, are {\it lower limits}.

Lower limits to the density of high redshift blazars powered by black holes with $M>M_9$
are placed by the existence of at least 4 blazars at $4<z<5$ for which G10 have estimated a
black hole mass larger than $10^9 M_\odot$. 
These blazars are 
RXJ 1028.6--0844 ($z=4.276$; Yuan et al. 2005);
GB 1508+5714 ($z=4.3$; Hook et al. 1995);
PMN J0525--3343 ($z=4.41$; Worsley et al. 2004a) and
GB 1428+4217 ($z=4.72$; Worsley et al. 2004b).
The lower limit in the 5--6 redshift  range corresponds to the existence of at least one blazar, 
Q0906+6930 at $z=5.47$, with an estimated
black hole mass of $2\times 10^9 M_\odot$ (Romani 2006).

These are all sources pointing at us.
The real number density of heavy black holes in jetted sources, $\Phi_{\rm RL}(z, M>M_9)$, 
must account for the much larger population of misaligned sources.
We have then multiplied the mass function of blazars by 
$2\Gamma^2=450$, i.e. we have assumed an average $\Gamma$--factor of 15,
appropriate for the BAT blazars analysed in G10.

Summarizing, the BAT blazar survey allowed to meaningfully construct the hard X--ray LF of blazars. 
G10 have also constructed the minimal evolution consistent with the existing data and the (few)
existing lower limits. 
At the high luminosity end the LF can be translated into the mass function of black holes with more
than one billion solar masses.
In Fig. \ref{data} we show $\Phi_{\rm RL}(z, M>M_9)$ as derived from the cosmological evolution 
model of A09 as a red stripe, and that derived from the ``minimal" LF as a green stripe.

\begin{table*} 
\centering
\begin{tabular}{lllllllll}
\hline
\hline
$z^a$     &All$^b$ &Radio loud  &Radio loud &$R>10$ &{\it Swift}/BAT$^f$  &Expected, $\Gamma=15^g$ &Expected, $\Gamma=5^h$ \\
        &        &$R>10$$^c$  &\%$^d$     &corrected$^e$     &  &  & 	   \\
\hline 
All     & 6194   & 576   & 9.30   &      &	         &  &\\           
1--2    & 1342   & 160   &11.92   &      &         &  &    \\ 
2--3    & 2541   & 260   &10.23   & 5200 & 20      &2000       &222\\ 
3--4    & 1706   & 129   & 7.56   & 1800 & 45      &4500       &500\\ 
4--5    & 550    & 21    & 3.81   & 252  & 52--78  &5200--7800 &580--870 \\
5--6    & 36     & 2     & 5.55   & 56   & 5--52   & 500--5200 & 55--580 \\ 
\hline
\hline 
\end{tabular}
\vskip 0.1 true cm
\caption{
$^a$ Redshift bin.
$^b$Number of uniformly selected quasars with $\log L_{\rm bol}>47$ (erg s$^{-1}$)
in the SDSS+FIRST survey (8770 square degrees in common).
$^c$ Number of objects with the radio (5 GHz) to optical (2500 \AA) monochromatic flux ratio larger than 10. 
$^d$ Column $c$ divided by column $b$.
$^e$ Number of radio loud objects expected in the SDSS+FIRST footprint, $\sim$8770 square degrees. 
The number is obtained from the SDSS 
QSO luminosity function of Hopkins et al. (2007) rescaled by the radio loud fraction given in column $d$. 
$^f$ The number of blazars detected by {\it Swift}/BAT with an estimated black hole mass exceeding 
$10^9 M_\odot$, from the minimal and the A09 evolutions, when relevant. 
$^g$ The expected number of radio--loud sources calculated by multiplying the number of detected blazars
(column $f$) in the given redshift bin by $450\times 8,770/40,000\sim 10^2$ (i.e. assuming $\Gamma=15$).
$^h$ Same as column $g$, but for $\Gamma=5$ (the number scales as $\Gamma^2$).
}\label{numbers}
\end{table*}

The BAT blazars described above can be compared
to the radio--loud sources in the quasar catalog of the 
Sloan Digital Sky Survey (SDSS, Schneider et al. 2010) 
Data Release Seven (DR7) that includes information on radio detection in
and the Faint Images of the Radio Sky at Twenty--cm survey 
(FIRST, Becker et al. 1995). 
The region of the sky covered by both surveys is $\sim$8770 square degrees.
We adopt the public catalog with quasar properties described in Shen et al. (2010),
which includes quasars bolometric luminosity (using bolometric corrections derived from 
the composite spectral energy distributions from Richards et al. 2006). 
The catalog also provides the radio flux density at rest--frame 6 cm and the optical flux density at rest--frame 
2500 \AA\ that can be used to calculate the radio--loudness. 
Following Jiang et al. (2007) we define
a source radio--loud if it has radio to optical flux ratio, $R$, larger than 10. For a handful of sources 
where optical quantities are not provided, we supplement the ``raw" catalog by calculating 
the bolometric luminosity from the absolute i--band magnitude, assuming a bolometric 
correction of 2.5. 
This bolometric correction is derived by matching the average bolometric 
luminosity provided by Shen et al. (2010) with the bolometric luminosity calculated from the absolute 
i--band magnitude for sources where the catalog lists both quantities. 
We then calculate the rest--frame optical flux from the luminosity, in order to derive an estimate of the 
radio--loudness.  This ``extended" catalog will be our reference.

We select all sources that are in the FIRST+SDSS footprint, have an optical bolometric luminosity 
$>10^{47}$ erg s$^{-1}$ and $R>10$. 
We also require the quasars to be selected uniformly using the final quasar target 
selection algorithm described in Richards et al. (2002). 
These amount to, e.g., 21 radio--loud sources in the $4<z<5$ redshift bin.
We note that this number {\it is not} representative of a complete, volume limited sample 
(i.e., of the true luminosity function). 
To derive a simple estimate of the statistical incompleteness, we compare the number of
quasars in Tab.~\ref{numbers} (column $b$) to the number predicted by the bolometric LF,
which is derived from surveys that include the SDSS (Hopkins et al. 2007, hereafter H07), 
in the SDSS+FIRST area.
For example, in the same redshift bins, approximately  a factor $\simeq 10$ more objects 
are expected from the LF proposed by H07.

On the other hand, such ``incompleteness" biases should not affect number ratios, such as 
the radio loud fraction (RLF) derived from our FIRST+SDSS sample. 
As a matter of fact, our derived RLF (column $d$ of Tab. \ref{numbers}) is  completely consistent 
with the  RLF derived by Jiang et al. (2007). 
This allows us to estimate the ``expected number'' of RL objects simply multiplying the values 
obtained from the H07  LF by the RLF we have derived from our sample. 
Such numbers are reported in column $e$ of Tab. \ref{numbers}, and should be compared to the 
expectations from detected BAT blazars (columns $g$ and $h$). 
It is immediately clear that, at least at $z\gsim 3$, the expectations largely exceed 
what derived from the observed LF if $\Gamma=15$, while $\Gamma=5$ seems to be 
quantitatively consistent with data.  For increasing redshift, the corrected fraction of RL objects (column $e$) 
is progressively lower, and its ratio with respect to columns $g$ and $h$ is decreasing (the figure in column $e$ exceeds both columns $g$ and $h$ 
for $z =2-3$,  it is in between the values of  columns $g$ and $h$ for $z = 3-4$, it is lower for $z = 4-5$ and $z = 5-6$. In the latter case, 
the corrected fraction of RL objects just coincides with the lower boundary of the range from column $h$, that adopts $\Gamma=5$). 
We note that a discrepancy between the observed number of radio-loud quasars and theoretical predictions was first noted 
by Haiman et al. (2004) and confirmed by McGreer et al. (2009). 

Before discussing the possible nature of such discrepancy, it is useful to analyze 
the predictions concerning the fraction of radio--loud sources at high redshifts.

\subsection{Radio--loud vs radio--quiet quasars at high redshift}

We now estimate the ratio of radio--loud to radio--quiet quasars having black holes exceeding
$M_9$ as a function of redshift.
We must therefore derive the number density $\Phi_{\rm RQ}(z, M>M_9)$ of radio--quiet quasars 
hosting black holes of $M>M_9$ in different redshift bins.
Consider then the bolometric LF of radio--quiet sources (H07)
and simply assume that quasars radiate at an average fraction $f_{\rm Edd}$ of the Eddington limit so that:
\beq
\frac{M}{10^9 {\rm M_\odot}}=3\times 10^{-14}\frac{1}{f_{\rm Edd}}\frac{L}{L_\odot}. 
\eeq
To derive $\Phi_{\rm RQ}(z, M>M_9)$ one then integrates the LF above the luminosity threshold, 
$L_{\rm min}$, corresponding to $M=M_9$.   

Estimates of  $\Phi_{\rm RQ}(z, M>M_9)$ for radio--quiet quasars are shown in 
Fig. \ref{data} for $f_{\rm Edd}=0.3$ ($L_{\rm min}=10^{13} L_\odot$) and $f_{\rm Edd}=1$ 
($L_{\rm min}=3.4\times 10^{13} L_\odot$).  
Note that the lower the average $f_{\rm Edd}$, 
the lower the luminosity of a quasar that hosts a $M>M_9$ black hole is. 
Therefore, decreasing $f_{\rm Edd}$ allows one to integrate
the LF down to lower luminosities, thus increasing the overall $\Phi_{\rm RQ}(z, M>M_9)$. 
However, if we were to assume, say, $f_{\rm Edd}=0.1$, then the range of luminosities 
where we compare radio--loud and radio--quiet quasars would differ, 
as in blazars the (observed) accretion disk component is always more luminous
than several $\times 10^{46}$ erg s$^{-1}$ (cf. Fig. \ref{blazars}).

Finally, if the SDSS misses quasars because of obscuration 
biasing optical selection (see below and Treister et al. 2011), then this mass function 
is in reality a lower limit, as more active black holes might exist.

\begin{figure}
\includegraphics[width=8.7cm]{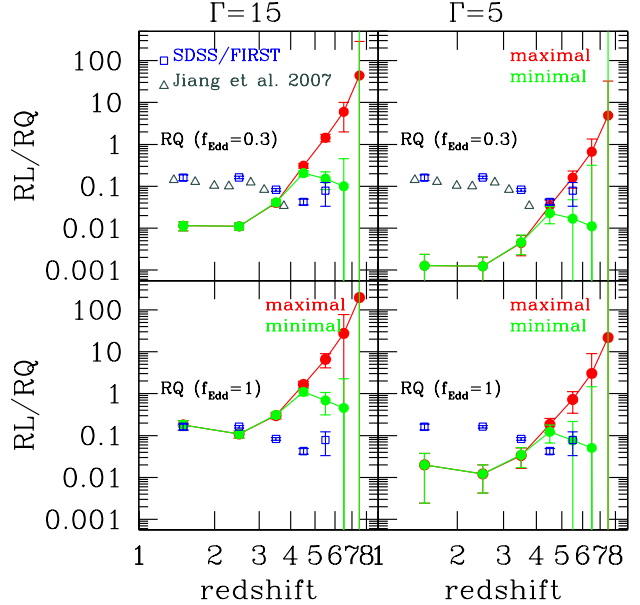}
\vskip -0.5 cm
\caption{
Ratio of the number density of radio--loud to radio-quiet quasars with black hole
masses exceeding $10^9 M_\odot$.
Blue and grey symbols refer to the ratio calculated for the SDSS sources (this work and Jiang et al. 2007 respectively),
while green (lower set) and red (upper set) dots refer to the ratio of the radio--loud population
inferred from the Fermi (at $z=1.5$) and BAT blazars and the radio--quiet quasars as derived from the SDSS
(H07).
Red points have been calculated assuming the A09 cosmic evolution of blazars;
green points correspond to the ``minimal" cosmic evolution of blazars suggested in G10. 
The radio--quiet population is estimated from the LF of radio--quiet quasars 
and its redshift evolution, from H07.
Top panel: average Eddington fraction of radio--quiet quasars is $f_{\rm Edd}=0.3$. 
Bottom panel: average Eddington fraction of radio-quiet quasars is $f_{\rm Edd}=1$. 
Left panels: $\Gamma=15$. 
Right panels: $\Gamma=5$.
} 
\label{RL}
\end{figure}

We quantify the redshift evolution of the ratio of radio--loud vs radio--quiet sources in Fig. \ref{data} 
and Fig. \ref{RL}.   
{\it We stress that up to $z=4$, where we do see blazars, the cosmological evolution model,  
as derived by A09, is secure}.  
Beyond $z=4$ it depends strongly on the assumed evolution. 
Since, however, the ``minimal" evolution provides a lower limit to the number of radio-loud systems, 
we can be assured that {\it the radio--loud vs radio--quiet fraction remains at least close to 
constant, and near unity, up until $z\simeq 6$}.  
We find that the fraction of jetted sources increases from $z=3.5$ to $z=4.5$ by roughly an order of magnitude. 
Fig. \ref{data} also  shows that for $M>M_9$  and $L>10^{13} L_\odot$ ($L \gsim 10^{47}$ erg s$^{-1}$) 
the number density of radio--loud quasars approaches and possibly prevails over that of radio quiet--quasars, 
if we take face value the extrapolation of the cosmic evolution suggested by A09.

We further check our results via a comparison of the radio--loud fraction that we derive from the 
FIRST+SDSS sample we uniformly selected. 
This radio--loud fraction is shown in Fig. \ref{RL} (blue squares). 
We compare our estimate with the results for quasars of similar luminosities of Jiang et al. 2007, 
shown as gray triangles (their Fig. 3, lower left panel).   The point at $z=1.5$ is based on the
Fermi blazars analyzed in Ghisellini et al. 2011, in order to obtain a radio--loud fraction at low redshift
and maximize the range where we can compare our results to Jiang et al. (2007).
The agreement between our selection in the SDSS+FIRST and Jiang's is excellent where the 
analyses overlap ($z \leqslant 4$). 
A striking result we find is that, while at $z \leqslant 2.5$ the 
``parent population" of SDSS+FIRST radio loud quasars traces almost perfectly the BAT 
blazars, assuming $\Gamma=15$ and $f_{\rm Edd}=1$, the two selections deviate at higher redshift.  

This analysis, which relies on a uniformly selected sample, does not require a volume complete
sample, as we are now dealing with fractions. 
We still find a dearth of radio--loud sources at high redshift. 

We stress once again that at $z<2.5$ the blazar population, with $\Gamma=15$, 
joined with the radio-quiet population with $f_{\rm Edd}=1$, is in excellent 
agreement with SDSS/FIRST data (both our analysis and Jiang et al. 2007 analysis). 
At $z=3.5$ the number density of blazars is derived from observed sources 
(no redshift extrapolation), and our only assumption is the value of $\Gamma$.  
We are therefore confident that there {\it must} be either a transition 
in the astrophysical properties of the population or a selection bias.
Below we discuss some possibilities to explain the found discrepancy.

\subsection{Where are radio--loud high--redshift quasars?}

The problem we face is simply that BAT blazars indicate that there must be 
many radio--loud sources that instead the SDSS+FIRST survey does not detect.
In the following we list possible solutions.

\begin{enumerate}

\item 
The value $\Gamma\sim 15$ for the average bulk Lorentz factor of blazars is too large. 
A value of $\Gamma\sim 5$ would make the predicted numbers of misaligned 
radio sources to decrease by an order of magnitude (it is proportional to $\Gamma^2$), 
becoming then consistent with the SDSS+FIRST detected radio--loud sources.
We have checked that fitting the SED of our blazars with $\Gamma=5$ 
gives reasonable results, but this value of $\Gamma$ cannot account for all the 
measured apparent superluminal velocity and, furthermore,
it implies that the jet is away from the conditions of minimum
jet power requirements (see Ghisellini \& Tavecchio 2010). 
Since the discrepancy appears as redshift increases,
we need that $\Gamma$ evolves with cosmic time ($\Gamma$ must decrease with 
increasing redshift).

A similar solution is that the jet has a radial velocity structure, similar to Gamma Ray Bursts,
thus emitting (at approximately the same level) within an angle larger than $1/\Gamma$.
We thus observe preferentially that part of the source exactly pointing
at us (i.e. with a viewing angle close to zero) maximally beamed.
The other slightly misaligned part of the source contribute less
to the observed flux.
This implies that we underestimate the jet power, that refers to the part
of the jet mostly contributing to the flux, the one pointing at us.
The other parts, viewed at angles larger than $1/\Gamma$, are not accounted
for when estimating the jet power.
In other words, we calculate the power of only a part of the jet, of solid angle
$\sim 1/\Gamma^2$, which is smaller than the jet solid angle $\Omega_{\rm j}\sim \theta_{\rm j}^2$,
where $\theta_{\rm j}$ is the 
jet half opening angle. 
Therefore,  very approximately, we underestimate the  jet power by the factor $(\theta_{\rm j}/\Gamma)^2$. 
To account for the disagreement in number, $(\theta_{\rm j}/\Gamma)^2$
should be one order of magnitude. 

\item 
There is a bias, in the SDSS+FIRST survey, against the detection of
powerful radio--loud sources at high--$z$, but not against radio--quiet quasars.
This (yet unknown) bias could be due to the compactness of the radio--halo
in radio--sources that are still too young to have developed an extended structure.
As a consequence, the flux emitted in this compact and isotropic structure
is self--absorbed up to the GHz frequency range.
Further study is however needed to verify this possibility, that now is only
a speculation.

\item The SDSS/FIRST selection misses a large fraction of powerful quasars
(both radio--loud and radio--quiet) at $z>3$. 
This may be the results of optical absorption, or else of collimation
of the optical emission of the disk 
(making the apparent disk luminosity much dimmer if the disk is observed edge on). 
An absorbed (or optically anisotropic) source intrinsically emitting 
more than $10^{47}$ erg s$^{-1}$ in the optical (our selection limit) would not pass
our selection.
This would occur also for radio--loud objects, because, even if the radio--emission is 
unaffected, the primary selection (SDSS) is on the optical luminosity.

The radio--loud fraction could be right (if the bias applies equally 
to both kind of sources), but both classes are under--represented
by an order of magnitude, at least at $z>3$.

\end{enumerate}

These possibilities are listed in order of increasing demands of heavy black holes
at large redshifts.
In fact, case i) would minimize the total number of high--$z$
powerful radio--loud sources (and their associated
heavy black hole), that would be well described by the SDSS+FIRST survey.
The price to pay is a correspondingly larger requirement on the jet power of each 
quasar (i.e. we have a factor 10 less radio--loud powerful sources, and then
less heavy black holes, but each jet is 10 times more powerful).
In this case the total number of heavy black holes at high--$z$ is practically
given by the radio--quiet quasars (the radio--loud contributing by less than 10\%).

Case ii) is intermediate, since it implies that the number of powerful radio--loud sources
is correctly described by the BAT blazars (with $\langle\Gamma\rangle\sim 15$), that should be
as numerous as the radio--quiet sources (with $M>M_9$) detected by the SDSS.
Therefore the number of heavy and early black holes is roughly twice as much as the one
derived by the radio--quiet quasars LF.
The most important consequence in this case is the strong evolution of the radio--loud
fraction, becoming close to unity at $z>3$, with interesting consequences on our understanding
of the growth of the early supermassive black holes.

Case iii) is the more demanding: it implies that the SDSS misses 90\% of the most powerful quasars
(no matter their radio--loudness) and thus that the number of heavy black holes is a whole 
order of magnitude more than the mass function derived from the H07 LF would predict.

In the following we will examine the current ideas about the relationship 
between dark matter halos and black hole mass, in order to estimate
the expected number of heavy black holes as a function of redshift.


\section{Black hole--dark halo connection}

Empirical correlations have been found between  
the central stellar velocity dispersion and the asymptotic circular 
velocity ($V_{\rm c}$) of galaxies (Ferrarese 2002, Baes et al. 2003, Pizzella et al. 2005):
\beq
\sigma=200 \kmps \left(\frac{V_{\rm c}}{320 \kmps}\right)^{1.35}
\eeq
and
\beq
\sigma=200 \kmps \left(\frac{V_{\rm c}}{339 \kmps}\right)^{1.04}
\eeq
as suggested by Pizzella et al. (2005) and Baes et al. (2003), respectively.
Some of these  relationships (Ferrarese 2002, Baes et al. 2003) mimic closely the simple $\sigma=V_c/\sqrt[]{3}$ definition that one derives
assuming complete orbital isotropy. We note that in an isothermal sphere $\sigma=V_c/\sqrt[]{2}$.

Since the asymptotic circular  velocity ($V_{\rm c}$) of galaxies is a measure of the 
total mass of the dark matter halo of  the host galaxies, one can relate in simple ways the mass 
of the central black hole to the mass of its host halo (``hole--halo" connection, e.g., 
Ferrarese 2002, Volonteri et al. 2003, Wyithe \& Loeb 2003, Rhook \& Wyithe 2005, Croton 2009). 
A halo of mass $M_{\rm h}$ collapsing at redshift $z$ has a circular velocity:
\beq V_{\rm c}= 142 \kmps \left[\frac{M_{\rm h}}{10^{12} \ M_{\sun} }\right]^{1/3} 
\left[\frac {\Omm}{\Ommz}\ \frac{\Delta_{\rm c}} {18\pi^2}\right]^{1/6} 
(1+z)^{1/2},  
\eeq 
where $\Delta_{\rm c}$ is the over--density at virialization relative 
to the critical density. 
For a WMAP5 cosmology we adopt here the  fitting
formula  $\Delta_{\rm c}=18\pi^2+82 d-39 d^2$ 
(Brian \& Norman 1998), 
where $d\equiv \Ommz-1$ is evaluated at the collapse redshift.
In this case we obtain 
\beq
\Ommz \, =\, { \Omm (1+z)^3  \over \Omm (1+z)^3+\Oml+\Omk (1+z)^2 }   
\eeq
We will further assume that the black hole--$\sigma$ ($M$--$\sigma$) scaling is:
\beq
\frac{M}{M_9} \, =\,  \left(\frac{\sigma}{356 \kmps} \right)^4.
\eeq
%
(Tremaine et al. 2002).
We also assume that these scaling relations observed in the local universe hold at all redshifts.  
Therefore we derive relationships between black hole and dark matter halo mass: 
%
\beq
\frac{M_{\rm h}}{10^{13} M_\odot}  \, =\, 4.1  \, (M/M_9)^{0.56} \, 
\left[ \frac{\Omm}{\Ommz} \frac{\Delta_c} {18\pi^2} \right]^{-1/2} 
(1+z) ^{-3/2} 
\eeq
if we adopt the relationship in Pizzella et al. (2005);
%
%
\beq
\frac{M_{\rm h}}{10^{13} M_\odot}\, =\, 8.2 \,(M/M_9)^{0.75} \,
\left[ \frac{\Omm}{\Ommz} \frac{\Delta_{\rm c}} {18\pi^2} \right]^{-1/2} 
(1+z)^ {-3/2} 
\eeq
if we adopt $\sigma=V_c/\sqrt[]{3}$ (almost equivalent to what one would calculate using 
the relationship in Baes et al. 2003).

\begin{figure}
\vskip -0.3 cm
\includegraphics[width=\columnwidth]{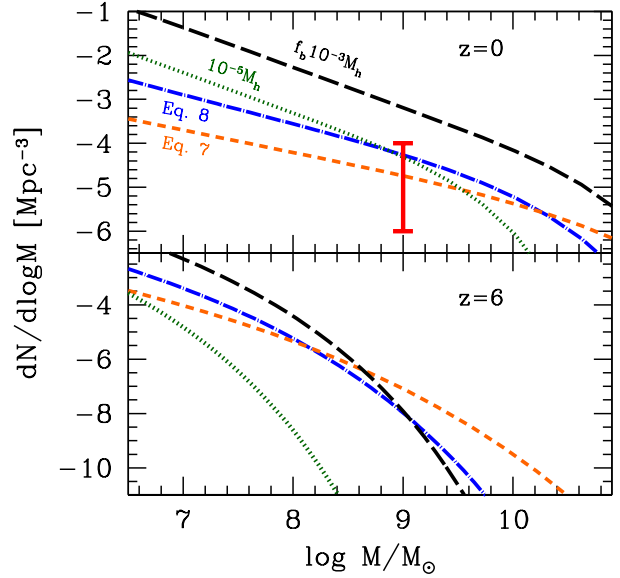}
\vskip -0.3 cm
\caption{
Mass functions of black holes at $z=0$ (top) and at $z=6$ (bottom). 
Vertical (red) bar: constraints at $z=0$ (G\"ultekin et al. 2009).
Dash--dotted (blue) curve: $\sigma=V_c/\sqrt[]{3}$ (Eq. 8)+ $M$--$\sigma$;
short dashed (orange) curve: Pizzella (Eq. 7) +$M$--$\sigma$.
As a reference, we also show also  the mass function obtained by assuming 
$M=f_{\rm bar}\,10^{-3} M_{\rm h}$(black long--dashed curve) 
and  the case $M=10^{-5} M_{\rm h}$
(green dotted curve).
}
\label{MF}
\end{figure}

\subsection{Black hole mass functions}

We can therefore estimate the mass function of black holes by convolving Eq. 7 and Eq. 8
with the mass density of dark matter halos with mass $M_{\rm h}$ derived from 
the Press \& Schechter formalism (Sheth et al. 1999). 
The number density of black holes with $M>M_9$, $\Phi(z, M>M_9)$, 
therefore corresponds to the number density of halos with mass  $M_{\rm h}>M_{\rm thr}$, if $M_{\rm thr}$ 
is the mass of a halo that hosts a billion solar masses black hole, and we assume that 
all halos host black holes.

\begin{figure}
\includegraphics[width= \columnwidth]{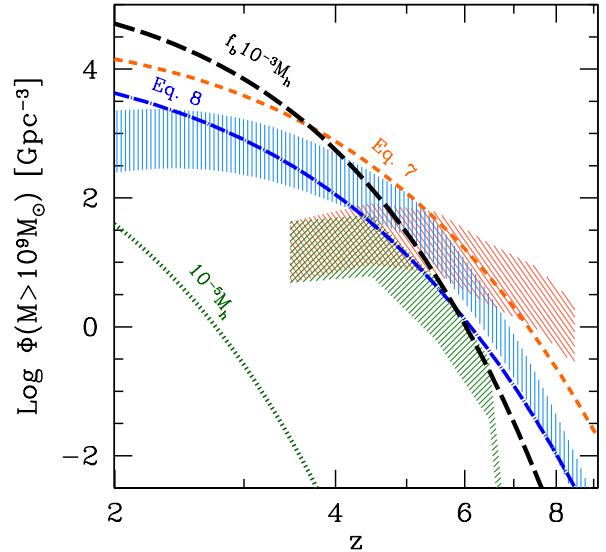}
\vskip -0.7 cm
\caption{
Number density of black holes with $M>10^9M_\odot$ as a function of redshift. 
Red hatched area: number density of black holes in radio--loud sources, derived from 
the blazar LF of A09, assuming $\Gamma=5$ (lower bound), 
or $\Gamma=15$ (upper bound). 
Green hatched area: ``minimal"  number density (studied in G10), assuming 
with $\Gamma=5$ (lower bound), or $\Gamma=15$ (upper bound).  
Blue  hatched area:  black hole number density of radio--quiet quasars (from
the LF and its evolution studied in H07) 
Line styles and colors  of the ``hole--halo'' connection predictions  as in Fig. \ref{MF}.
}
\label{n_z6}
\end{figure}

In Fig. \ref{MF} we show the mass functions derived via Eq. 7 and Eq. 8
coupled to the Press \& Schechter function at $z=0$ (bottom panel) and $z=6$ (top panel). 
As an exercise, one can also assume that the black hole mass scales linearly with the halo mass. 
We can derive a plausible upper limit to this scaling assuming $M=f_{\rm b} 10^{-3} M_{\rm h}$, 
where $f_{\rm b}=\Omega_{\rm b}/\Omega_{\rm M}\simeq 0.14$ is the universal baryon fraction, 
and $M\simeq10^{-3} M_{\rm bulge}$ is the empirical correlation between black hole and bulge 
mass in elliptical galaxies (Marconi \& Hunt 2003, Haring \& Rix 2004). This assumption corresponds 
to assuming that galaxies do not loose any baryon because of feedback effects, and all baryons 
end up in a stellar bulge.  At $z=0$ this relationship is obviously wrong (the baryon content 
in galaxies is much less than $f_{\rm b}$ and not all baryons end up in stellar bulges), 
and we find that $M=10^{-5} M_{\rm h}$ provides a more acceptable solution, as shown 
in the upper panel of Figure~\ref{MF}.  At $z=0$ the mass function of black holes is 
estimated in the literature by coupling the empirical correlations 
found between black hole mass and host properties (bulge mass, luminosity and velocity dispersion, 
see Marconi et al. 2004, Gultekin et al. 2009 and references therein) with the distribution of 
galaxies as a function of these properties.
In the top panel of figure~\ref{MF} we show with a vertical bar the current limits on
the mass density of black holes with $M>M_9$ at $z=0$. 
The ``hole--halo''  connection coupled to the Press \& Schechter function might possibly 
be slightly overestimating the number density of large black holes at $z=0$.  
This is due to the LF of galaxies being steeper at the high mass end than the 
halo mass function at the high mass end. 

We stress here that at least two biases do exist that affect the comparison between the 
``hole--halo" connection and observational samples: first, Lauer et al. (2007)
suggest that the $M-\sigma$ relation compared to $M-L$  relation {\it underestimate} 
the number of BH with $M>M_9$ (but see Bernardi et al. 2007; Tundo et al. 2007). 
Second, going in the same direction, the intrinsic scatter in the $M-\sigma$ relation allows a  
larger number of possible haloes hosting massive black holes   
(see Lauer et al. 2007b; Gultekin et al. 2009).  
We discuss the importance of the intrinsic scatter in \S \ref{scattersection}.

Finally, Kormendy et al. (2011a) question any correlation between 
black holes and dark matter halos (but see Volonteri et al. 2011).
We notice that Kormendy's argument is not a concern here, as at large masses 
Kormendy et al (2011b) suggest that a `cosmic conspiracy' causes $\sigma$ and $V_c$ to 
correlate, thus making the link between $M$ and $V_c$ adequate. 
Although estimates we derive from the halo--hole connection 
are therefore {\it extremely} uncertain, they can still provide some sense of 
the possible hosts of these massive black holes at high redshift.

\subsection{Number density of high-redshift $M>10^9 M_\odot$ black holes powering blazars}

We now turn to compare the  number density of $M>M_9$  and 
$L>10^{13} L_\odot$ ($L \gsim 10^{47}$ erg s$^{-1}$) blazars 
to the upper limits defined by ``hole--halo" connection.  
The various assumptions for the ``hole--halo" connection discussed in \S 3 
are shown  in Fig. \ref{n_z6}.  
The number density of heavy black holes powering jetted sources  
is now close to or even greater (if we extend the cosmological  
evolution model of A09 beyond z$\sim$4) than the upper limit defined by ``hole- halo" connections 
 at the largest redshifts. The mass function derived by the ``minimal" LF is instead consistent.

\begin{figure}
\vskip -0.3 cm
\includegraphics[width=\columnwidth]{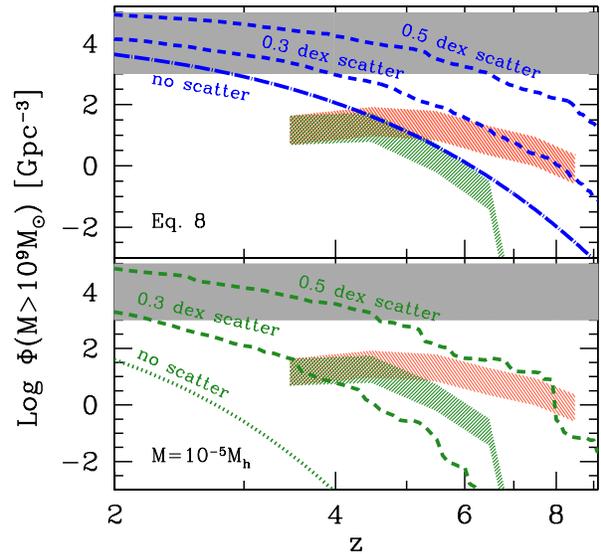}
\vskip -0.6 cm
\caption{
Importance of scatter for the prediction of the number density of black holes
with $M>10^9M_\odot$ as a function of redshift. 
Top panel: the $\sigma=V_c/\sqrt[]{3}$ (Eq. 8)+$M-\sigma$ curve (blue dashed dotted)
increases by assuming a scatter of 0.3 and 0.5 dex (dashed blue curves).
Bottom panel: even the $M=10^{-5} M_{\rm h}$ curve (green dotted) can become consistent
with the LF of radio--loud sources (hatched areas) assuming a scatter of 0.5 dex.
The grey shaded area in both panels indicate the limits on today's number density 
of $M>M_9$ black holes.
}
\label{scatter}
\end{figure}

Despite the uncertainties, we can make some simple inferences on the ``hole--halo" connection at high-redshift: 
the number density of $M>M_9$ black holes cannot be below the limit imposed by the number 
density of blazars derived from the ``minimal" evolution. For instance, it can be ruled 
out that at high redshift the black hole mass scales as $M=10^{-5} M_{\rm h}$, if scatter is negligible. 
This implies that high redshift black holes represent a higher fraction of the mass of 
a dark matter halo (at least, this is the case for the most massive active black holes, 
$M>M_9$ and $L>10^{13} L_\odot$, this might not be the case for lower mass holes, 
see, e.g. the discussion in Willott et al. 2010).   
Including the effect of scatter eases such constraints, as we show below. 

{\subsection{Importance of scatter}
\label{scattersection}

We can assume that
at fixed $\sigma$ the logarithmic scatter in black hole mass is $\Delta=$0.3-0.5 dex
($M_{\rm BH}=M_{\rm BH,\sigma}\times 10^{\Delta \delta}$, where $\delta$ 
is normally distributed, see, e.g., Gultekin et al. 2009).
We include scatter, at various levels of $\Delta$, by performing a Monte Carlo simulation, where for each
black hole mass we create  500 realizations of the host mass.  
Fig. \ref{scatter} shows two examples of Monte Carlo realizations of the mass function and number density 
that include scatter at the  level of $\Delta=$0.3--0.5 dex (Merloni et al. 2010).
By comparing Fig. \ref{scatter} to Fig. \ref{n_z6} one clearly sees how 
scatter dramatically increases the number of 
black holes with  $M>M_9$, and for most ``hole--halo" connection the number density 
of high--redshift radio--loud sources can be accommodated. 
For instance, the $M-\sigma$ relations can accommodate the radio--loud population 
as long as scatter is around 0.3 dex.  
 Notice that in the logarithmic scale of Fig. \ref{scatter} today's number density 
of $M>M_9$ black holes (red vertical bar in the upper panel of Fig.4) is between 3 and 5 (grey stripe in Fig. \ref{scatter}), 
comparable to the  number density at $z\simeq5$ in the cases with significant scatter.x For instance, if we assume the scaling of Eq. 8, the number density of  $M>M_9$ black holes reaches 
$10^3$ Gpc$^{-3}$ at $z=4$ for  $\Delta=$0.3 and $z=7$ for  $\Delta=$0.5. If $\Delta=$0.5 the upper limit to today's number density,
$10^5$ Gpc$^{-3}$, is reached at $z=2$, implying that after that cosmic time  the number density of $M>M_9$ black holes cannot 
increase any more.

We have hitherto assumed that the duty cycle, $x_{\rm dc}=1$, corresponding to the fraction of 
black holes that are active (related to the ratio of the lifetime of quasars to the Hubble time),
that is, all halos host an active black hole. We have further assumed that 
radio--loud quasars dominate at high redshift.  
If the active fraction $x_{\rm dc}$ is less than unity, implying that not all black holes with  
$M>M_9$ are active and accreting close to the Eddington rate, than the 
requirements become stricter. 
Similarly, if radio--quiet quasars dominate the population of active 
black holes, their number has to be accounted for as well. In the next section we expand our 
analysis to include radio-quiet sources and allow for an active fraction, or duty cycle, below unity.

\begin{figure}
\includegraphics[width= \columnwidth]{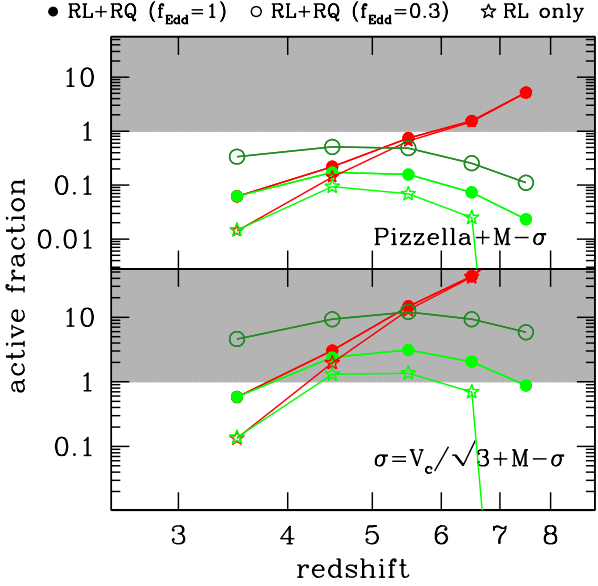}
\vskip -0.3 cm
\caption{
Fraction of $M>10^9 M_\odot$  that is active.
Red: A09 cosmic evolution inferred from the LF of blazars (this curve is basically
unaffected by different choices of $f_{\rm Edd}$ as radio-loud quasars dominate the 
active population). 
Green: ``minimal" cosmic evolution inferred from the LF of blazars. 
Stars: radio--loud systems only. 
Filled circles: all active quasars, where we assumed $f_{\rm Edd}=1$. 
Empty circles: all active quasars, where we assumed $f_{\rm Edd}=0.3$. 
Bottom:  $\sigma=V_c/\sqrt[]{3}$ +  $M-\sigma$. 
Top:  Pizzella +  $M-\sigma$. 
In all cases we ignore scatter in the $M-\sigma$ relation. 
The grey shaded area is not permitted as the active fraction becomes larger than unity. 
} 
\label{AF}
\end{figure}

\subsection{Active fraction of high redshift $M>10^9 M_\odot$ black holes}

Until now we have focused only on the constraints that high-redshift blazars impose on
the number density of $M>M_9$ black holes. However, this is clearly a lower limit to the number
of massive black holes that have to exist at high redshift, as we have to take into account 
radio--quiet sources,
for instance including optically selected quasars as described in Section 2.1. 
If the bolometric LF of radio--quiet sources  
(H07) is a good tracer of the quasar population, than there are 
roughly 0.1 radio-loud quasars per each radio--quiet one (assuming the minimal evolution 
of the LF of blazars), up to one or more, if we take the evolution of A09 face value. 

However, based on the arguments of \S 2.1 and \S 2.2, there might be a large population of
radio-quiet quasars that are not accounted for in the bolometric LF of radio--quiet sources  
of H07 (see case iii in \S 2.2). Since we have no information on this 
putative hidden population we do not here include this speculation in our 
analysis, but if in reality most quasars are missed by current surveys, then all 
problems we discuss below are exacerbated. 

We can assess the requirements on the population of high--redshift BH hosts by investigating 
the fraction of  $M>M_9$  black holes that are actively accreting, that is, 
the active fraction of black holes. 
We define the active fraction as:
\beq
x_{\rm dc}({\rm RL})=\frac{\Phi_{\rm RL}(z, M>M_9)}{\Phi(z, M>M_9)}
\eeq
for radio-loud blazars only, and
\beq
x_{\rm dc}{\rm (RL+RQ)}=\frac{\Phi_{\rm RL}(z, M>M_9)+\Phi_{\rm RQ}(z, M>M_9)}{\Phi(z, M>M_9)}
\eeq
for all active sources, where $\Phi_{\rm RL}(z, M>M_9)$ and $\Phi_{\rm RQ}(z, M>M_9)$ 
are defined in section 2.   
$\Phi_{\rm RQ}(z, M>M_9)$ depends on one parameter, 
the typical Eddington ratio of radio--quiet black holes.  
$\Phi(z, M>M_9)$ is defined in \S 3.1, and depends on the ``hole-halo" connection 
and on the level of scatter that this relationship suffers. 
The numerator and denominator in both Eq.~9 and~10 are derived independently, hence
a priori the active fraction can apparently assume values above unity. 
When $x_{\rm dc}>1$ the result is however unphysical, and it allows us to rule out 
a given model. 

The active fraction is shown in Fig. \ref{AF} for $f_{\rm Edd}=1$,  $f_{\rm Edd}=0.3$ and 
negligible scatter in the ``hole-halo" connection. 
We here show both the total active fraction, $x_{\rm dc}$(RL+RQ), and the active fraction in 
radio--loud sources only, $x_{\rm dc}$(RL). 
In both cases we have assumed $\Gamma=15$.  
Face value, by $z\simeq 5$ almost all ``hole--halo" connections, except for the case of Eq. 7,  
require an active fraction (or a duty cycle) of unity.  
A significant amount of scatter might however alleviate the issue, as 
shown in Fig. \ref{AF_scatter}. 
As discussed above, scatter increases very significantly the number density in $M>M_9$ 
black holes, and pushes the active fraction to much lower values, at the cost however 
of having already built up almost all $M>M_9$  black holes by $z\simeq 5$ (cf. Fig.~6).

Finally, if black holes accreted at rates significantly below the Eddington rate, the estimate of the 
number density of radio--quiet quasars that we derive from the luminosity function
would increase (\S 3).  This decreases the radio-loud fraction (see Fig. 5), but at the same 
time the {\it total} active fraction increases, and the active fraction becomes close to 
unity even for a significant scatter of 0.3 dex.

\begin{figure}
\includegraphics[width= \columnwidth]{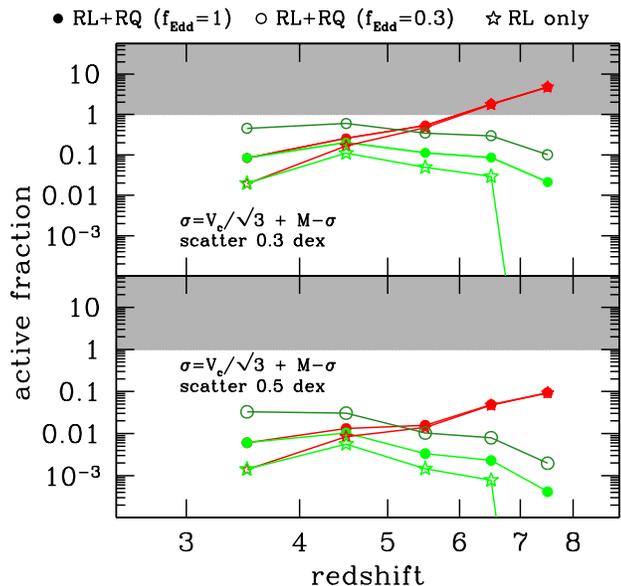}
\vskip -0.3 cm
\caption{
Same as Fig. \ref{AF}, for the case $\sigma=V_c/\sqrt[]{3}$ +  $M-\sigma$, but including scatter.
}
\label{AF_scatter}
\end{figure}

\section{Discussion and conclusions}

We investigated the relative occurrence of radio--loud and radio--quiet quasars
in the first billion years of the Universe, based on the sample of high--redshift 
blazars detected in the 3--years all sky survey performed by 
the Burst Alert Telescope (BAT) onboard the {\it Swift} satellite (Ajello et al. 2009).
The masses $M$ of the black holes powering these quasars exceeds a few billions 
solar masses, with accretion luminosities being a large fraction of the Eddington limit (G10).
For each blazar pointing at us, there must be hundreds of similar sources (having 
black holes of similar masses) pointing elsewhere.  
This can set constraints on the number density of dark matter halos that can host 
massive black holes at high redshift. 

We first compared the number of radio--sources hosting heavy black holes
estimated from the BAT detected blazars to the SDSS+FIRST survey 
to explore the relative importance of (jetted) radio--loud vs radio--quiet sources. 
We find a rough agreement up to $z\sim 3$, but beyond this redshift
there is a deficit of radio sources (detected by the FIRST and present in the SDSS surveys)
with respect to the expectations (see also Haiman et al. 2004, McGreer et al. 2009).
We found no obvious explanation for this deficit, and have suggested three
possibilities: i) the bulk Lorentz factor of the jet (controlling
the number of misaligned sources) decreases beyond $z=3$ (from $\Gamma \sim 15$ to $\Gamma\sim 5$);
ii) there is a bias against detecting distant (and therefore possibly young)
radio--sources with the FIRST survey (i.e. at 1.4 GHz) and iii) there is a bias against optical 
selection of distant and powerful quasars, both radio--quiet and radio--loud,
due to absorption or collimation of the disk emission.
These possibilities affect our estimates of the number density of heavy 
black holes ($M>10^9 M_\odot$) in an increasing way [from possibility i) to iii)].

In the first case the majority of quasars are radio--quiet at all redshifts, and the 
number density of high--redshift $M>10^9 M_\odot$ black holes can be safely 
derived from the observed LF of radio--quiet quasars.
In the  second case radio--loud and radio--quiet quasars powered by heavy black holes
become comparable in number beyond $z\sim$3--4, doubling the number
of heavy and high--$z$ black holes estimated from radio--quiet quasars only.
This would also mean that the radio--loud fraction increases with $z$ for
sources with heavy black holes, suggesting that that a radio phase is perhaps
a necessary ingredient for fast black hole growth at early cosmic times.
The last possibility is the most demanding, since it implies that 
we see only a minor fraction of the intrinsically luminous high--$z$ quasars
(both radio--loud and quiet), implying that although the radio--loud fraction
is always of the order of 10\% (i.e. at all redshifts), the number of
heavy black holes is now severely under--estimated (by one order of magnitude).

We re--iterate that there is a good agreement between the number density of $M>10^9M_\odot$ 
black holes found with blazars and the total number of radio--loud quasars up to $z\sim 3$,
but not beyond.
We then conclude that even if we do not know the cause for this disagreement, there must be
some change occurring at $z\sim 3$.

We then studied plausible ``hole--halo" connections in order to predict
the number of supermassive black hole at high redshifts.
We found that, unfortunately, the predicting power of these
relations is weak, mainly because of the large effect 
that scatter can have: since we are dealing with fastly
declining functions (corresponding to the high end of the distributions of
luminosities and/or black hole masses), it is possible that even a few
large black hole inhabiting halo slightly less massive than 
implied by the adopted relation can dominate the number density
at a given redshift.

On the other hand, despite the rather large uncertainties,
the sources that have been already observed (cf. \S~2) suggest that 
large and distant black holes are all active (or nearly so),
and that they are all Eddington limited (or nearly so).
If not, the number of heavy black holes would be larger, and the simple
theoretical ideas we have adopted in this paper 
would start to have some difficulties to account for them.

Finding more luminous radio--quiet quasars at $z>5$ is obviously mandatory to 
study the high mass end of the black hole mass density,
but we would like to stress that finding high redshift blazars might be, in the end,
even more important, since each one of those implies the existence of many more
misaligned sources.
A few blazars detected at $z\sim 6$ would be very challenging for
structure formation, very constraining, and possibly illuminating for 
the understanding the early growth of very massive black holes,
and its feedback on the host.
The existence of these blazars possibly implies that normal ``feedback" 
might not be at play at the highest redshifts. 
A possible explanation is 
that high--accretion rate events, distinctively possible during the violent 
early cosmic times, trigger the formation of collimated outflows 
that do not cause feedback directly on the host, which is pierced through. 
These jets will instead deposit their kinetic energy at large distances, 
leaving the host unscathed. This is likely if at large accretion rates photon 
trapping decreases the disk luminosity, while concurrently the presence of a 
jet helps dissipating angular momentum, thus promoting efficient accretion.  
This picture may explain why high--redshift massive black holes can accrete at very high rates 
without triggering self--regulation mechanisms.

\section*{Acknowledgments}
We thank M.~J.~Rees for enlightening discussions.  MV acknowledges support from SAO Awards TM9-0006X, TM1-12007X and NASA award ATP NNX10AC84G.  RDC acknowledge financial support from ASI (grant n.I/088/06/0).

\bibliographystyle{mn2e}

\label{lastpage}

\end{document}